# On Manipulation in Prediction Markets When Participants Influence Outcomes Directly

MITHUN CHAKRABORTY and SANMAY DAS, Washington University in St. Louis

## 1. INTRODUCTION

Prediction markets are platforms designed to aggregate dispersed information about some future event from a potentially diverse crowd of informants. It is generally assumed that the agents who participate in the market by trading may have superior information about the relevant event, but have no direct control over the outcome. However, prediction markets are often used in situations where this assumption is violated to a greater or lesser degree. In fact, sometimes it is the very potential for violation that gives agents the informational edge that prediction markets get their value from.

Consider three of the canonical real-world examples where prediction markets (or betting markets) have demonstrated their forecasting ability to great effect: elections/politics [Berg et al. 2008], sporting events [Wolfers and Zitzewitz 2004], and software product releases [Cowgill et al. 2010]. In each of these cases, it is easy to see how the presence of a prediction market on the event may distort incentives. For instance, a congressional staffer or member of congress may know more about the probable outcome of a key vote than the general public, but is also in a position to influence said outcome.

While it is acknowledged that prediction markets have value as forecasting tools that may help in making business and policy decisions, they have gone through cycles of hype and bust for reasons that include regulatory concerns about manipulation. The emblematic anecdote about this problem is the failure of DARPA's proposed Policy Analysis Markets which were caricatured in the media as "terrorism futures" [Hanson 2007b; Stiglitz 2003]. There are obviously prediction markets that will not work but stock and futures markets have been used for a long time as predictive tools, and prediction markets are no different in essence. The key is to understand when markets may be prone to manipulation and how much to trust them. There is a large body of work, both theoretical and experimental, on various types of manipulative behavior in prediction markets, their impact, and possible remedies [Ottaviani and Sørensen 2007; Chen et al. 2009; Shi et al. 2009; Hanson et al. 2006].

We introduce a new game-theoretic model of prediction markets that captures two aspects of real-world prediction markets: (1) agents directly influence the outcome that the market is set up to predict, (2) some of the agents who have influence over the outcome may not participate in the prediction market (*e.g.* not every voter trades in an election prediction market). We are not concerned with markets where an individual has a very small effect on the outcome (like large elections), so we focus initially on a two-player model, which helps us understand "worst case" manipulative behavior. Among other things, we study what effect, if any, the uncertainty around the participation of some outcome-deciders has on the actions of a strategic agent when there is incentive for manipulation.

Our model is a two-stage game. In the first stage, the two players have the opportunity to participate (sequentially) in a prediction market mediated by a *market scoring rule* or MSR [Hanson 2007a] market-maker. In the second stage, the two players take public actions which we term "votes" for convenience, although in general they model each participant's role in determining the outcome. The payoffs from the stage one prediction market are determined by the stage two vote. While the model is simple, the analysis yields several interesting theoretical insights.

## 2. MODEL

Let $\tau \in T$ denote the unknown type of the "entity" on which the combined market and voting mechanism is predicated (*e.g.* for the vote-share prediction market of a two-candidate election, the "entity" would be one of the two candidates). The two agents, whom we call Alice and Bob ($A$ and $B$ in subscripts) following convention [Chen et al. 2009], receive private signals $s_A, s_B \in \Omega = \{0, 1\}$. The prior distribution $\Pr(\tau)$ on the type and the conditional joint distribution $\Pr(s_A, s_B|\tau)$ on the private signals given the type are common knowledge. Let $q_0(s) \triangleq \Pr(s_B = 0|s_A = s)$, $s \in \{0, 1\}$ denote Alice's posterior probability that Bob received the signal $s_B = 0$, given her own signal and common knowledge.

In the market stage (the starting market price is assumed to be $p_0 = 1/2$), Alice trades first with an MSR market maker and changes the price to $p_A$. Then, it is Bob's turn to trade but he may not show up with a probability $\pi$ (known to Alice) which is called Bob's *non-participation probability*; if he does trade, he changes the price to $p_B$. In the voting stage, Alice and Bob simultaneously announce signals $v_A, v_B$ respectively from $\Omega = \{0, 1\}$. The *liquidation value* of the market security is the average vote $v = (v_A + v_B)/2 \in \{0, \frac{1}{2}, 1\}$. [1] If Bob did not trade in the first stage, we assume that he votes truthfully, declaring his private signal. Any agent participating in the market is Bayesian and strategic, and their payoff is given by $r_i(p_i, p_j, v_A, v_B) = s(p_i, (v_A + v_B)/2) - s(p_j, (v_A + v_B)/2)$, $i \in \{A, B\}$, where $s(\cdot, \cdot)$ is the strictly proper scoring rule used to implement the market; $j = 0$ for $i = A$, $j = A$ for $i = B$.

## 3. EQUILIBRIUM ANALYSIS

LEMMA 3.1. *If Alice updates the market price to $p_A > p_0 = 1/2$ (resp. $p_A < 1/2$), then her rational vote is $v_A = 1$ (resp. $v_A = 0$), regardless of Bob's vote.*

LEMMA 3.2. *Under mild assumptions, there exist unique values $p^L \in (0, 1/2)$ and $p^H \in (1/2, 1)$ such that if $p_A > 1/2$ (resp. $p_A < 1/2$), then Bob's rational vote is $v_B = 1$ if $p_A \leq p^H$ also (resp. if $p_A < p^L$ also), and is $v_B = 0$ otherwise. Table I gives the values of $p^L, p^H$ for three representative scoring rules.*

|       | Logarithmic MSR | Quadratic MSR | Spherical MSR |
|-------|-----------------|---------------|---------------|
| $p^L$ | 0.2             | 0.25          | 0.2725        |
| $p^H$ | 0.8             | 0.75          | 0.7275        |

Table I.

For example, if Bob does trade and observes $p_A \in (1/2, p^H)$, then his rational action is to drive $p_B$ all the way to $1$ and vote $v_B = 1$, but if $p_A \in (p^H, 1)$, then he should set $p_B = 1/2$ and vote $v_B = 0$.

With the help of the above lemmas, we can show that for every $(\pi, q_0)$ pair, the perfect Bayesian equilibrium of this game belongs to one of two possible categories, depending on Bob's participation probability. Below a threshold on Bob's probability of participation in the trading stage, say $(1 - \pi_c)$, we call the equilibrium a low participation probability (LPP) equilibrium, and above $(1 - \pi_c)$, we call it a high participation probability (HPP) equilibrium. The *cross-over* probability $\pi_c$ is a function of the MSR used and the value of $q_0$ (Alice's posterior on Bob's signal), as shown in Figure 1.

**Main result:** In a LPP equilibrium, Alice predicts Bob's vote, and then bases her trading on the optimal combination of her own vote and Bob's vote, and the prediction market price is reflective of the expected outcome. In a HPP (collusive) equilibrium, on the contrary, Alice expects Bob to enter and collude with her, and she chooses a prediction market price that allows Bob and her to "split the profit" (not necessarily evenly); specifically, she sets $p_A = p^H$ if $q_0 < 1/2$ and $p_A = p^L$ if $q_0 > 1/2$. We can analyze the characteristics of these equilibria from the following two perspectives.

---

[1] In general, we can have $v = \alpha v_A + (1-\alpha)v_B$, $\alpha \in (0, 1)$, where $\alpha$ models the degree of control that Alice alone exerts over the final outcome. Here, we focus on the special case $\alpha = \frac{1}{2}$ as a starting point where both agents are equally powerful.

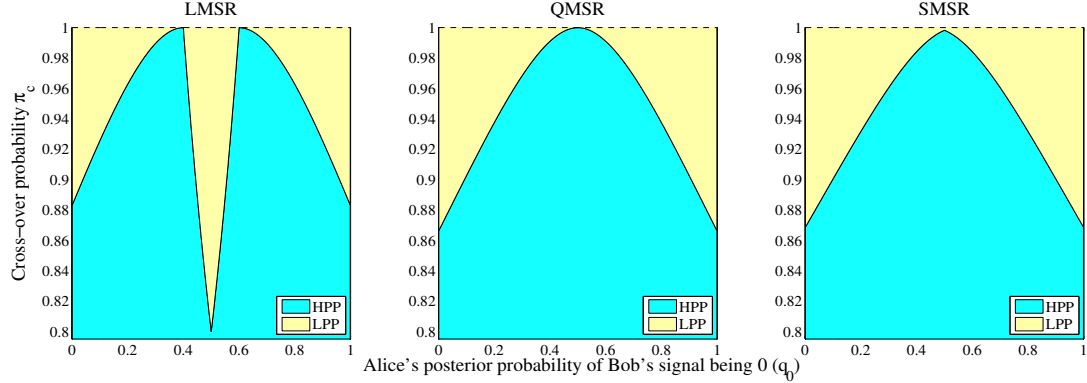

Fig. 1: Left to right: the dependence of the cross-over probability (Bob's non-participation probably that serves as the threshold between the two types of equilibria) on Alice's posterior probability of Bob having a signal $0$ for the logarithmic (LMSR), quadratic (QMSR), and spherical (SMSR) scoring rules. For example, for QMSR, if $q_0 = 0.25$, then $\pi_c \approx 0.9537$, so we have a LPP equilibrium with $p_A = (1 + \pi(1 - q_0))/2$ for $\pi > 0.9537$, and a HPP equilibrium with $p_A = p^H = 0.75$ for $\pi < 0.9537$. Note the peculiar form of the graph for LMSR due to the properties of the logarithmic scoring function.

### 3.1 Informativeness of prices

We can examine the information on the outcome conveyed by prices after each stage of trading.

|  | **Bob trades** | | **Bob does not trade** | |
| --- | --- | --- | --- | --- |
|  | **LPP** | **HPP** | **LPP** | **HPP** |
| $p_A$ | Bayesian estimate | Predetermined, $\in \{p^L, p^H\}$ | Bayesian estimate | Predetermined, $\in \{p^L, p^H\}$ |
| $p_B$ | $v$ (actual outcome) | $v$ (actual outcome) | Bayesian estimate | Predetermined, $\in \{p^L, p^H\}$ |

Here "Bayesian estimate" means that $p_A^{LPP} = \mathbb{E}\left[v|\pi, q_0, p_A = p_A^{LPP}\right]$, that is, Alice's posterior expectation of the average vote conditional on the information available at that time to everyone but Bob, and hence an effective disseminator of information (note that if Bob does not trade, the final price $p_B = p_A$).

### 3.2 Truthfulness of actions

It is easily seen that the trading actions of either agent do not necessarily reveal their private signals. Regardless of whether the equilibrium is LPP or HPP, if Bob does end up participating in the trading stage, his vote is fully determined already by Alice's trading choice, independent of his signal.

Alice's trading decision, on the other hand, is based on her belief about Bob's signal which, in our Bayesian setting, has an indirect dependence on her own information. In fact, under the mild assumption of *stochastic relevance* [Miller et al. 2005] of Alice's signal for Bob's signal, *i.e.* if $q_0(s_A = 0) \neq q_0(s_A = 1)$, it is possible to recover Alice's true signal $s_A$ from her price-update $p_A$ in an LPP equilibrium since $p_A^{LPP}$ is a simple function of $\pi$ (known) and $q_0$. This stands in contrast to the situation where Bob's participation is certain, an extreme case of the HPP domain. In an HPP equilibrium, $p_A$ is merely a (known) constant independent of the actual values of the belief parameters; the only information we can extract from it is whether $q_0 > 1/2$ (for $p_A = p^L$) or $q_0 < 1/2$ (for $p_A = p^H$), which is insufficient for deducing $s_A$ uniquely in the absence of further assumptions on the belief structure. Thus, for a sufficiently high probability that the successor (Bob) will not trade and will vote truthfully (*i.e.* $\pi > \pi_c$), the predecessor, though strategic, is forced to act in a way that divulges her private information indirectly!